\documentclass[twocolumn,showpacs,preprintnumbers,amsmath,amssymb]{revtex4}
\usepackage{graphics}
\usepackage{epsfig}
\usepackage{bm}
\usepackage{umlaute}

\begin{document}
\title{Temperature and magnetic field dependences of the elastic constants of Ni-Mn-Al magnetic Heusler alloys}

\author{Xavier Moya, Llu\'{i}s Ma\~nosa}
\email{lluis@ecm.ub.es}
\author{Antoni Planes}

\affiliation{Departament d´Estructura i Constituents de la
Mat\`eria, Facultat de F\'isica, Universitat de Barcelona,
Diagonal 647, E-08028 Barcelona, Catalonia, Spain}

\author{Thorsten Krenke and Mehmet Acet}
\affiliation{Fachbereich Physik, Experimentalphysik,
Universit\"{a}t Duisburg-Essen, D-47048 Duisburg, Germany}

\author{Michel Morin}
\affiliation{Groupe d'Etudes de M\'{e}tallurgiePhysique et
Physique des Mat\'{e}riaux, INSA, 20, Av. A. Einstein, 69621
Villeurbanne, France}

\author{J.L. Zarestky and T.A. Lograsso}
\affiliation{Ames Laboratory, Department of Physics, Iowa State
University, Ames, Iowa 50011}
\date{\today}

\begin{abstract}

We report on measurements of the adiabatic second order elastic
constants of the off-stoichiometric Ni$_{54}$Mn$_{23}$Al$_{23}$
single crystalline Heusler alloy. The variation in the temperature
dependence of the elastic constants has been investigated across
the magnetic transition and over a broad temperature range. Anomalies
in the temperature behaviour of the elastic constants have been
found in the vicinity of the magnetic phase transition. Measurements
under applied magnetic field, both isothermal and variable temperature,
show that the value of the elastic constants depends on magnetic order,
thus giving evidence for magnetoelastic coupling in this alloy system.

\end{abstract}


\maketitle

\section{\label{sec:level1}Introduction}

Among bcc-based solids undergoing martensitic transformations,
magnetic shape memory alloys have received much attention in
recent years due to their distinct properties arising from the
coupling between structure and magnetism \cite{Magnetism05}. This
coupling, which results in magnetic field induced strains, makes
these materials candidates for potential use in device
applications. This effect occurs in ferromagnetic systems
undergoing a martensitic transformation and leads to the magnetic
shape memory (MSM) effect. It originates from the reorientation of
martensite domains and/or from the magnetic field induced
austenite/martensite transformation \cite{OHandley98}. Intimately
related to magnetic shape memory is the magnetocaloric effect
(MCE) \cite{Hu00,Marcos02}. The fact that in some alloys the
magnetization of the martensitic phase is lower than that in the
parent phase leads to an inverse magnetocaloric effect
\cite{KrenkeNat05}.

The prototype MSM system Ni-Mn-Ga has been most extensively studied
because of the presence of magnetic field induced strains as large as 10\%
\cite{Ullakko1996,Sozinov2002}. From a fundamental point of view,
this alloy exhibits singular lattice-dynamical behaviour.
Specifically, it has been shown that the transverse TA$_2$ branch
shows a dip (anomalous phonon) at a particular wavenumber, at which
the energy softens with decreasing the temperature
\cite{Zheludev95,Stuhr97,Manosa01}. The temperature dependence of
the energy of the anomalous phonon parallels that of the shear
elastic constants, which also soften with decreasing temperature
\cite{Worgull96,Manosa97,Stipcich04}. Significant magnetoelastic
coupling exists in this system as evidenced by
the enhancement of the anomalous phonon softening, when the sample
orders ferromagnetically \cite{Planes97,Stuhr97,Manosa01}, and by
the change in the elastic constants when a magnetic field is applied at
constant temperature \cite{GonzalezComas99}.

The brittleness of the Ni-Mn-Ga compound has prompted the search
for MSM material with mechanically more favourable properties such
as Ni-Mn-Al. Stoichiometric Ni$_2$MnAl is structurally stable down
to low temperatures, but martensitic transformations occur within
a certain range of compositions close to stoichiometry
\cite{Morito98}. The low temperature martensitic structure depends
on composition, and the observed structures are the same as those
reported for other Ni-Mn based magnetic shape memory alloys
\cite{Morito98,Morito96,Pons00,Krenke05,Krenke06}. Previous
studies have revealed that at low temperatures, Ni$_2$MnAl
consists of a mixed L2$_1$+B2 phase, which incorporates
ferromagnetic and antiferromagnetic parts
\cite{Acet02,Manosa03,Rhee04}. The magnetic ordering in the
metastable B2 phase of Ni$_2$MnAl is conical antiferromagnetic
\cite{Ziebeck75}, and the presence of the B2 phase in this mixed
state is due to the low B2$-$L2$_1$ order-disorder transition as
compared to that of Ni-Mn-Ga. This leads to slow kinetics for the
ordering process \cite{Manosa03}.

Recently, we have carried out elastic and inelastic neutron
scattering experiments in a Ni-Mn-Al single crystal, for which we observe
significant TA$_2$ phonon softening at $\xi_{0} = 0.33$ \cite{Moya06}. In
addition, a number of elastic satellites associated
with magnetic ordering and with structural instabilities were also
observed. It has been shown that these precursor satellites only
grow once the sample orders magnetically suggesting that
there is an interplay of magnetism and structure in Ni-Mn-Al at a microscopic
level.

The aim of the present work is to extend the study of the lattice
dynamics of the Ni-Mn-Al system to the long wavelength limit by
measuring the elastic constants. We pay special attention to the
effect of applied magnetic field and provide
information on the interplay between structure and magnetism in
the long wavelength limit.

\section{Experimental Details}

The sample investigated is a single crystal with composition
Ni$_{54}$Mn$_{23}$Al$_{23}$ grown by the Bridgman method; the
estimated error in the composition is $\pm 1$ at. \%. Details of
sample preparation are given in reference \cite{Moya06}. The
single crystal used in the present study was cut from the larger
single crystal used in previous neutron diffraction experiments at
the High Flux Isotope Reactor (HFIR) at the Oak Ridge National
Laboratory (ORNL) \cite{Moya06}. It was oriented using the X-ray
Laue backscattering technique. From the original rod, a
parallelepiped specimen with dimensions $8.70 \times 10.25 \times
10.75$ mm$^{3}$ with faces parallel to the $(1 \overline{1} 0)$,
$(110)$ and $(001)$ planes, respectively, was cut using a low
speed diamond saw. A thinner sample (2.45 mm thick) with faces
parallel to the $(1\overline{1}0)$ plane was also cut from the
original rod in order to measure slow modes, which may be affected
by strong attenuation. In addition, small oriented pieces cut from
the ingot close to the extracted parallelepiped specimen were cut
for magnetization measurements.

The velocity of ultrasonic waves was determined by the pulse-echo
technique. $X$-cut and $Y$-cut transducers with resonant
frequencies of 10 MHz were used to generate and detect the
ultrasonic waves. The transducers were acoustically coupled to the
surface of the sample by means of Nonaq stopcock grease in the
temperature range $10-250$ K, by Dow Corning Resin 276$-$V9 in the
temperature range $210-320$ K, and by Crystalbond509 (Aremco
Products, Inc.) in the temperature range $310-390$ K.

A standard closed-cycle helium refrigerator (displex) was used for
measurements below room temperature and up to 300 K. The
temperature inside the displex was measured with a silicon diode.
For high temperature measurements (up to 390 K) the sample was
placed into a copper sample holder which was heated by means of a
heating plate. In this case, the temperature was
measured with a Pt-100 resistor embedded into the sample holder in
close proximity to the sample.

A device has been constructed that allows both isofield and
isothermal measurements of the ultrasonic velocities. The sample
and the transducer are placed into a copper sample holder
equipped with a Peltier element. The sample is placed on the upper
surface of the Peltier element, of which the bottom surface is in good
thermal contact with the copper sample holder. The sample holder
is placed between the poles of an electromagnet and lies on top of a
copper cylinder. The copper cylinder is partially immersed into a bath of ice
and water. The temperature is measured by
a thermocouple attached to the sample. By controlling the current
input into the Peltier element it is possible to achieve a fine
tuning of temperature within the range $250-310$ K. Temperature
oscillations are less than 0.1 K. The gap between the magnetic
poles is 28 mm which enables fields up to 13 kOe to be applied
perpendicular to the propagation direction of the ultrasonic
waves.

The magnetization was measured as a function of temperature ($4-300$ K)
and field (up to 5 T) using a superconducting quantum interference device magnetometer.

\section{Experimental results and discussion}

\begin{figure*}
\includegraphics[width=0.7\linewidth,clip=]{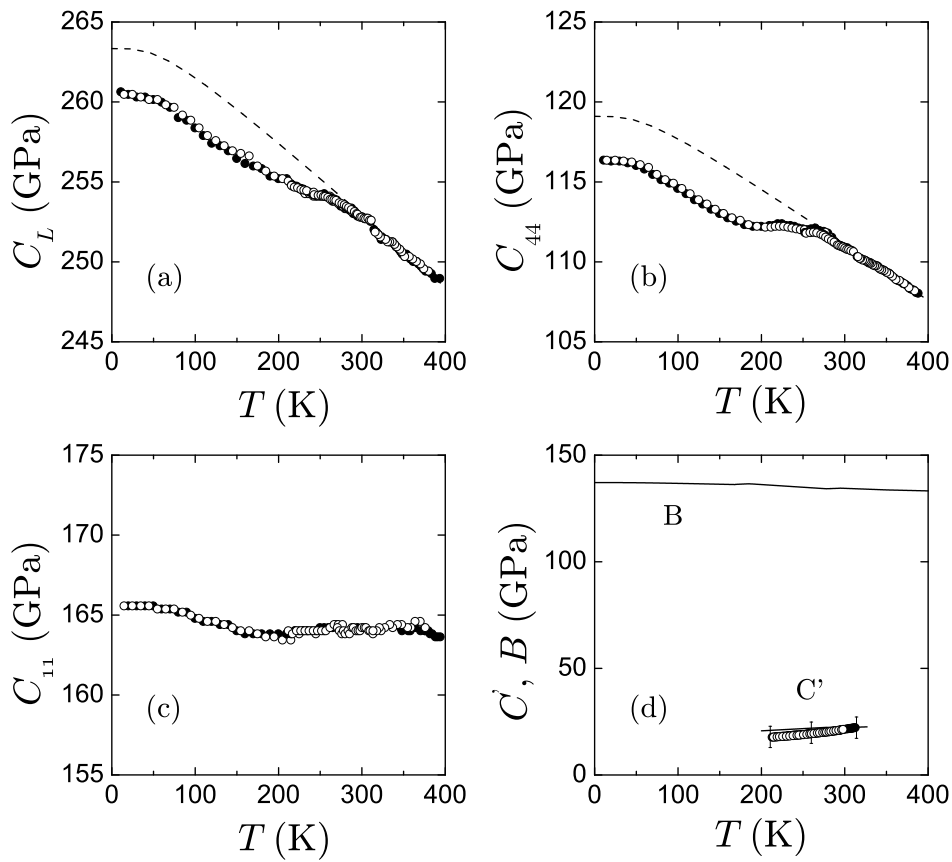}
\caption{Temperature dependence of the elastic moduli (a) $C_{L}$,
(b) $C_{44}$, and (c) $C_{11}$. Cooling and heating runs are
represented by open and closed symbols respectively. The dashed
lines show the expected behaviour for a Debye solid. Figure
\ref{fig1}(d) shows the values of the bulk modulus ($B$) and the
shear elastic constant ($C^{'}$) calculated from the complete set
of elastic constants measured (solid lines). The values of $C^{'}$
measured from the thinner sample are also shown for comparison.}
\label{fig1}
\end{figure*}

\subsection{Temperature dependence of the elastic constants}

For the parallelepiped sample, we have measured the velocity of
the longitudinal and shear waves propagating along the $[001]$
direction. Along the $[110]$ direction, we have also measured
longitudinal and $[001]$ polarized waves. Due to strong
attenuation it has not been possible to obtain reliable echoes for
the $[1\overline{1}0]$ polarized shear waves. For such  a mode, it
was possible to obtain echoes on the thinner sample. The values of
the elastic constants at room temperature derived from these
measurements are $C_{L}= 253 \pm 3$ GPa, $C_{44}= 111 \pm 2$ GPa,
$C_{11}= 164 \pm 3$ GPa, and $C^{'}= 21 \pm 4$ GPa. It is worth
noting that the value of $C^{'}$ computed using the data obtained
on the large crystal $C^{'}=C_{44}+C_{11}-C_{L}=22 \pm 8$ GPa is
in good agreement with the value obtained for the smaller sample.
Moreover, present values are in good agreement with those obtained
from the initial slopes of the acoustic phonon branches recently
measured by inelastic phonon scattering \cite{Moya06}.

From the temperature dependence of ultrasonic waves along the
$[110]$ and $[001]$ directions, we have computed the temperature
dependence of three independent elastic constants $C_{L}$,
$C_{44}$, and $C_{11}$. These are shown in figures \ref{fig1}(a),
(b), and (c) respectively. The data correspond to cooling (open
circles) and heating (closed circles) runs, and are obtained as an
average over several independent runs. From previous neutron
scattering measurements, the change of the cubic unit cell
parameter between 10 K and 350 K was estimated to be about 0.3 \%.
Therefore, no corrections for temperature induced changes of the
samples dimensions were taken into account. All elastic constants
increase with decreasing temperature and saturate at low
temperature. It should be noted that $C_{11}$ is almost
independent of temperature. This behaviour is similar to that
reported for Ni-Mn-Ga \cite{Stipcich04}. Dashed lines in figures
\ref{fig1}(a) and (b) represent the expected temperature behaviour
described by the quasiharmonic approximation of lattice dynamics
using the Debye model \cite{Lakkad71}. Within this framework, the
temperature dependence of the elastic constants is given by

\begin{equation}
C_{ij}=C_{ij}^{0}[1-\kappa F(T/\theta_{D})] \, ,
\end{equation}
where
\begin{equation}
F(T/\theta_{D})=3{(T/\theta_{D})}^{4}
\int_{0}^{T/\theta_{D}}\frac{x^{3}}{e^{x}-1} dx \, .
\end{equation}
$C_{ij}^{0}$ is the elastic constant at zero temperature,
$\theta_{D}$ the Debye temperature, and $\kappa$ a material
dependent parameter. It is seen that the measured elastic
constants show an anomalous temperature dependence deviating from
the Debye behaviour within the temperature range $200-300$ K,
approximately.

Figure \ref{fig1}(d) shows the temperature dependence of the bulk
modulus ($B$) and the shear elastic constant ($C^{'}$) calculated
from the complete set of measured elastic constants (solid lines).
The values of $C^{'}$ obtained from the ultrasonic velocity
measured on the thinner sample are also shown for comparison. We note
that the strong attenuation of the shear waves
associated with this mode makes it difficult to perform an
accurate measurement of this elastic constant, and the values are
affected by considerable error as shown by the error bars in
fig. \ref{fig1}(d). It was also not possible to obtain reliable
echoes below 200 K. The measured and calculated values are in good
agreement within experimental error. It should be noted that
$C^{'}$ exhibits a low value and softens with decreasing
temperature. Such a temperature softening is typical for bcc-based
solids which undergo martensitic transformations \cite{Planes01}
and reflects the dynamical instability of the cubic lattice
against shearing of $(110)$ planes along the $(1\overline{1}0)$
direction. It is worth noting that the elastic anisotropy
$A={C_{44}}/{C^{'}}=5.3$ is significantly lower than for Cu-based
prototypical shape-memory alloys \cite{Planes96}.

Previous studies on polycrystalline Ni-Mn-Al samples with
composition close to the studied sample revealed a magnetic
transition occurring around 300 K \cite{Manosa03}. This temperature
corresponds to the start of the deviation of the measured
elastic constants from Debye behaviour. This suggests that the
observed deviation from Debye behaviour may be due to the
development of magnetic order. In addition, neutron scattering
experiments also revealed the development of magnetic ordering
below $ \sim 300$ K for the crystal used in the present experiments.

Recent \textit{ab initio} calculations for stoichiometric
Ni$_2$MnAl have reported a value of the bulk modulus $B=157$ GPa
and $B=141$ GPa for ferromagnetic and antiferromagnetic ordering
respectively \cite{Busgen04}. The value derived from
the present ultrasonic measurements at $T=10$ K is in good agreement
with the computed value for antiferromagnetic order.

\begin{figure}
\includegraphics[width=0.7\linewidth,clip=]{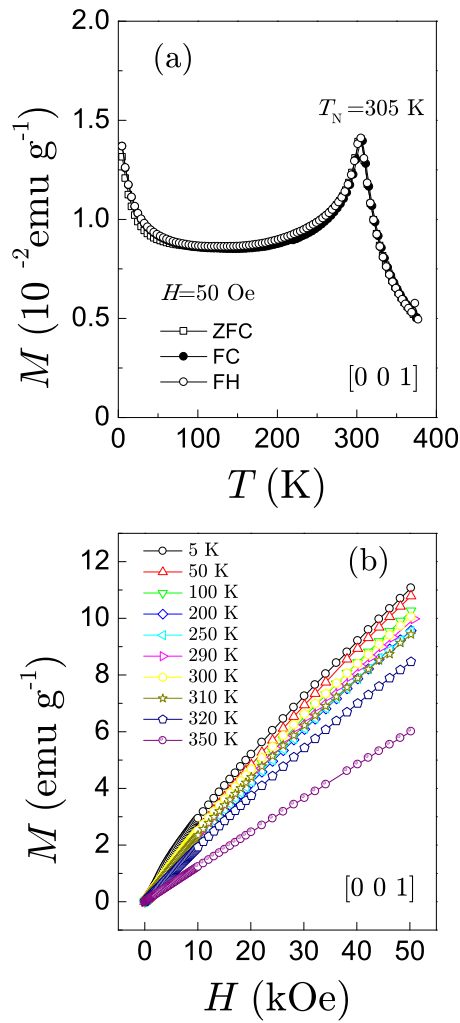}
\caption{(a) ZFC, FC, and FH $M(T)$ in a field of $H$=50 Oe
applied along the $[001]$ crystallographic direction. (b)
Magnetization versus field at several temperatures above and below
the magnetic transition. Prior to each $M(H)$ measurement, the
samples were prepared in the ZFC state by bringing them above 380
K. Solid lines are guides to the eye.} \label{fig2}
\end{figure}

\begin{figure*}
\includegraphics[width=0.9\linewidth,clip=]{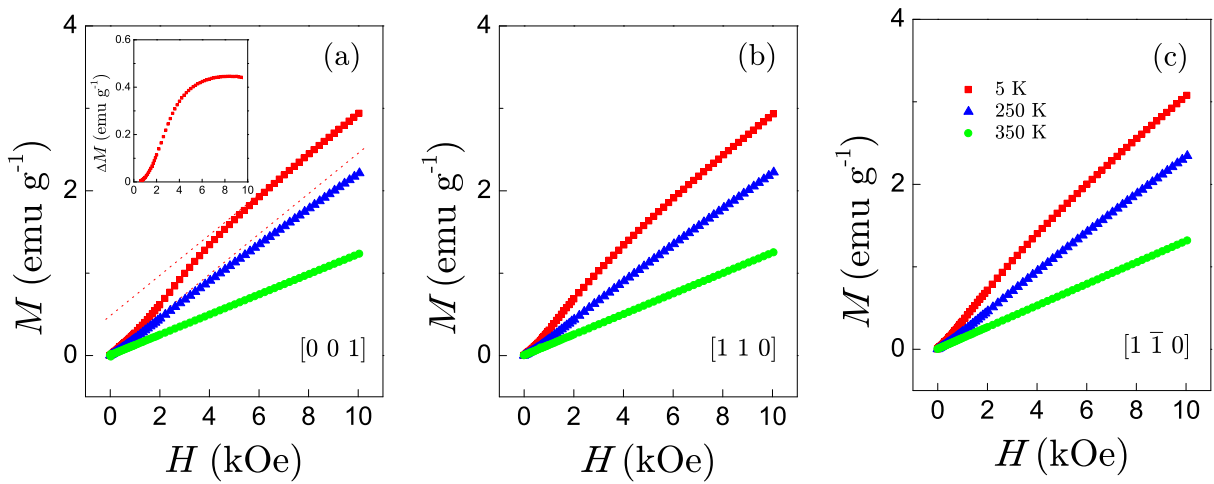}
\caption{(Color online). Magnetization versus field curves for
small fields along three crystallographic directions (a) $[001]$,
(b) $[110]$, and (c) $[1\overline{1}0]$ at selected temperatures.
The dashed lines in figure \ref{fig3}(a) represent the linear
behaviour extrapolated at small and high fields at 5 K. The inset
shows the deviation of the magnetization from the linear behaviour
corresponding to the small field region.} \label{fig3}
\end{figure*}

\subsection{Magnetization measurements}

With the aim of characterizing the magnetic behaviour of the
alloy, we have performed magnetization measurements as a function
of the temperature and magnetic field. Figure \ref{fig2}(a) shows
the temperature dependence of the magnetization $M(T)$ in a low
external magnetic field ($H=50$ Oe) applied along the $[001]$
crystallographic direction in the temperature range 5 K $\leq
T\leq$ 380 K. Prior to the measurements, the sample was prepared
in a zero-field-cooled state (ZFC) by cooling it from 380 K to 5 K
in the absence of a magnetic field. Subsequently, the external
field was applied and the measurements were taken on increasing
temperature up to 380 K. Then, without removing the external
field, the measurement was made on decreasing temperature, i.e.,
field-cooled (FC). As a last step, again without removing the
external field, the magnetization was measured on increasing
temperature. The last step is denoted as the field-heated (FH)
sequence. FC and FH curves retrace each other showing that no
structural transition occurs in the studied sample. The
magnetization exhibits a peak at 305 K which corresponds to the
magnetic transition from the paramagnetic state to the mixed
ferro/antiferromagnetic state \cite{Acet02}. This temperature
agrees well with the temperature below which magnetic satellites
have been observed by neutron diffraction. Due to the absence of
any considerable splitting between ZFC and FC curves together with
the low value of the magnetization, it is expected that the
ordering in the sample is mainly antiferromagnetic. This behaviour
is in agreement with previous studies on polycrystalline samples
with composition close to the studied sample \cite{Manosa03}. The
upturn of the magnetization below 130 K could be related to a
reorientation of the conical structure, as was stated in reference
\cite{Acet02}.

To further characterize the magnetic properties, the magnetic
field dependence of the magnetization $M(H)$ was measured in
fields up to 50 kOe applied along the $[001]$, $[110]$, and
$[1\overline{1}0]$ directions. In figure \ref{fig2}(b), we
illustrate the behaviour obtained at selected temperatures for the
field in the $[001]$ direction. The magnetization versus field
data at 350 K, which correspond to the paramagnetic state, show
typical linear behaviour of a paramagnet. Similar behaviour
is observed for the $M(H)$ data at 320 K. However, in this case
the data deviate slightly from linearity because of the presence
of short range FM correlations close to the transition temperature. Below
the magnetic transition, the $M(H)$ data continue to show nearly linear
behaviour, as expected for an antiferromagnet. Although the
magnetic state of the sample is mainly antiferromagnetic, the
deviation from linearity at low fields indicates that some
ferromagnetic order is also present in the system. It is worth
mentioning that the $M(T)$ and $M(H)$ behaviour measured in the
$[110]$ and $[1\overline{1}0]$ directions behave similar to that
reported for the $[001]$ direction.

The detailed view (low fields) of the magnetization versus field
curves at the selected temperatures is shown in figure \ref{fig3},
along the three crystallographic directions (a) $[001]$, (b)
$[110]$, and (c) $[1\overline{1}0]$. The magnetization data at 350
K show the typical linear behaviour for a paramagnetic state. By
contrast, below the magnetic transition, the $M(H)$ curves deviate
from the linearity at small fields but recover the linear
behaviour at fields higher than 5 kOe. Such a behaviour is
illustrated in figure \ref{fig4}(a) where the dashed lines
represent the linear behaviour extrapolated at small and high
fields at 5 K. The inset shows the deviation of the magnetization
from the linear behaviour. It is seen that the deviation saturates
at about 5 kOe. This behaviour indicates the presence of some
degree of ferromagnetic order, which is in agreement with previous
magnetic characterization of Ni-Mn-Al
\cite{Acet02,Manosa03,Rhee04}.

By comparing the temperature dependence of the elastic constants
[figure \ref{fig1}(a) and (b)] and magnetization [figure
\ref{fig2}(a)] it is seen that the deviation from the Debye
behaviour occurs within the temperature range of the magnetic
transition. In the vicinity of the magnetic transition, the
elastic constants are influenced by the onset of the magnetic
order showing a reduction of $C_{L}$ and $C_{44}$. The magnetic
contribution to the elastic constants, obtained as the difference
between the normal Debye behaviour extrapolated from the
paramagnetic region and the measured value, is negative for both
modes. This behaviour is in contrast with that displayed by
Ni-Mn-Ga ferromagnetic system which in the absence of magnetic
field, shows no significant change in any elastic moduli at the
Curie point \cite{Stipcich04}. We note that deviations from Debye
behaviour have been observed in antiferromagnetic systems such as
chromium \cite{Muir87} and Mn-Invar alloys
\cite{Cankurtaran93,Kawald94} below their Neel temperatures.

\subsection{Magnetic field dependence of the elastic constants}

\begin{figure*}
\includegraphics[width=1.0\linewidth,clip=]{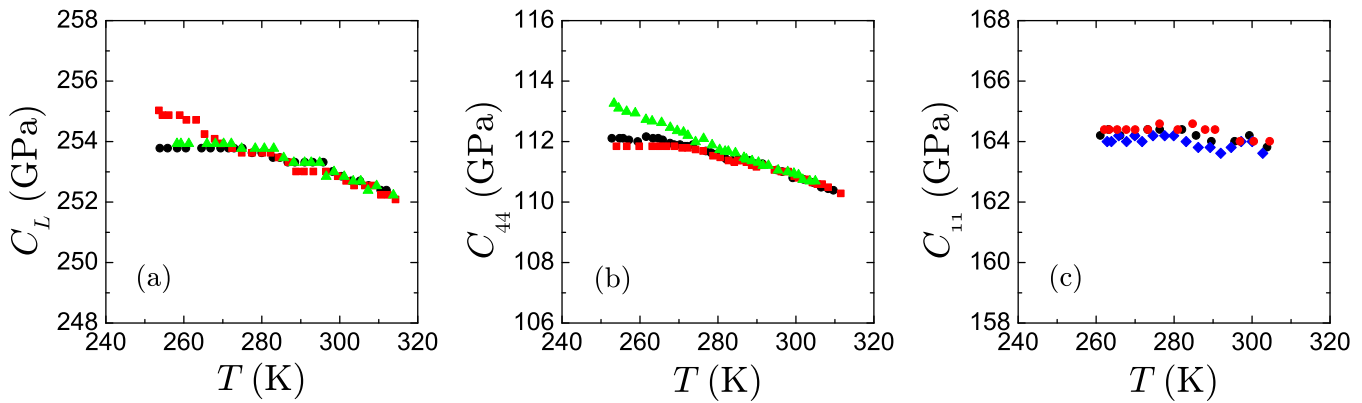}
\caption{(Color online). Temperature dependence of the elastic
constants measured while cooling across the magnetic transition
under magnetic field $H=5$ kOe (a) $C_{L}$, (b) $C_{44}$, and (c)
$C_{11}$. Circles represent the values without applied magnetic
field. Square, triangle, and diamond symbols represent
measurements obtained with magnetic field applied along the
$[1\overline{1}0]$, $[001]$, and $[110]$ direction respectively.}
\label{fig4}
\end{figure*}

\begin{figure*}
\includegraphics[width=1.0\linewidth,clip=]{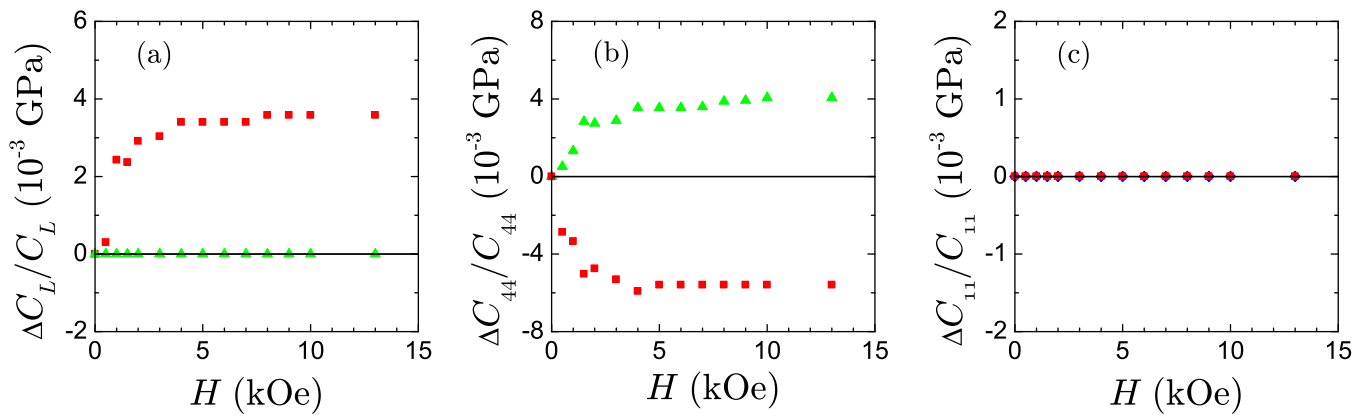}
\caption{(Color online). Magnetic field dependence of the relative
change of the elastic constants with respect to the value at zero
field at $T=263$ K. Square, triangle, and diamond symbols
represent measurements obtained with magnetic field applied along
the $[1\overline{1}0]$, $[001]$, and $[110]$ directions
respectively.} \label{fig5}
\end{figure*}

To investigate further the interplay between elastic and magnetic
properties, we have measured the elastic constants under isofield
and isothermal conditions. Application of magnetic field caused
strong distortion of the ultrasonic echoes of the $[110]$
$[1\overline{1}0]$ mode, and, consequently, it was not possible to
measure the magnetic field dependence of $C^{'}$. Figure
\ref{fig4} shows the temperature dependence of the elastic
constants measured under an applied magnetic field of $H=5$ kOe
along different crystallographic directions perpendicular to the
propagation direction of the ultrasonic waves. The data correspond
to cooling runs through the magnetic transition, both without
magnetic field and with applied field. Figures \ref{fig4}(a) and
\ref{fig4}(b) show respectively the values of $C_{L}$ and $C_{44}$
measured along the $[110]$ propagation direction (circles). Square
symbols represent measurements obtained with the magnetic field
applied along the $[1\overline{1}0]$ direction and triangle
symbols represent those obtained with the magnetic field along the
$[001]$ direction. Figure \ref{fig4}(c) shows the values of
$C_{11}$ measured along the $[001]$ propagation direction
(circles). The diamond symbols correspond to measurements obtained
with the field along the $[110]$ direction. From the figure, it is
seen that the three elastic constants behave differently when a
magnetic field is applied. While $C_{11}$ is not affected by the
magnetic field, $C_{L}$ and $C_{44}$ exhibit different behaviour
with and without applied field. On the one hand, the value of
$C_{L}$ increases for field along the $[1\overline{1}0]$
direction, while it is not affected by a field along the $[001]$
direction. On the other hand, $C_{44}$ decreases for a field along
the $[1\overline{1}0]$ direction and increases for a field along
the $[001]$ direction. The different behaviour exhibited by the
elastic constants evidences an anisotropic magnetoelastic coupling
for Ni-Mn-Al.

The behaviour found for the temperature dependence of the elastic
constants under magnetic field is confirmed by complementary
isothermal measurements. The relative change of the elastic
constants with respect to the value at zero field at $T=263$ K is
presented in figure \ref{fig5}. $C_{11}$ is not affected by
magnetic field, while $C_{L}$ and $C_{44}$ change when a field is
applied. This is consistent with the measurements performed on
cooling through the magnetic transition temperature with applied
magnetic field; $C_{L}$ increases up to saturation for a field
along the $[1\overline{1}0]$ direction, while it is not affected
by a field along the $[001]$ direction; $C_{44}$ decreases down to
saturation for a field along the $[1\overline{1}0]$ direction and
increases up to a saturation for a field along the $[001]$
direction. In the ferromagnetic Ni-Mn-Ga system, all elastic
constants increase up to saturation value with increasing magnetic
field \cite{GonzalezComas99}. Note that all relative changes
saturate at a field of approximately 5 kOe. Comparison with the
behaviour of the magnetization at small fields (figure
\ref{fig3}), shows that the behaviour of the relative change of
the elastic constants is related to non-linearities observed in
low fields in the magnetization versus field curves. These also
saturate at 5 kOe.

\section{Summary and conclusions}

Ultrasonic methods have been used to determine the elastic
constants of a Ni-Mn-Al Heusler alloy over a broad temperature
range. The room temperature values are in excellent agreement with
the values obtained from the slopes of the phonon dispersion
curves. The low temperature bulk modulus agrees well with the
reported value for antiferromagnetic Ni$_2$MnAl from
\textit{ab-initio} calculations. This gives experimental
confirmation for the softening to volume changes in the
antiferromagnetic state with respect to the harder ferromagnetic
state. A deviation from Debye behaviour has been observed below
the magnetic transition for all modes. Such a softening of the
lattice is a result of magnetoelastic coupling. Elastic constant
measurements under magnetic field indicate the existence of an
anisotropic magnetoelastic coupling. Such a coupling saturates at
about 5 kOe for all modes, and is associated with the non-linear
behaviour at low fields exhibited by the magnetic field dependence
of the magnetization.

\begin{acknowledgments}

We acknowledge Sophie Gu\'{e}zo for her collaboration. This work
received financial support from the CICyT (Spain), Project No.
MAT2004--01291, Marie--Curie RTN MULTIMAT (EU), Contract No.
MRTN--CT--2004--505226, DURSI (Catalonia), Project No.
2005SGR00969, and from the Deutsche Forschungsgemeinschaft
(GK277). XM acknowledges support from DGICyT (Spain). Ames
Laboratory is operated for the U.S. Department of Energy by Iowa
State University under Contract No. W-7405- Eng-82. The work at
Ames was supported by the Director for Energy Research, Office of
Basic Energy Sciences.

\end{acknowledgments}


\bibliography{apssamp}

\end{document}